\documentclass[journal]{IEEEtran}

\normalsize

\ifCLASSINFOpdf

\else

\fi

\usepackage{cite}

\usepackage{graphicx}

\usepackage{psfrag}

\usepackage{subfigure}

\usepackage{url}

\usepackage{stfloats}

\usepackage{amsmath}
\usepackage{array}

\usepackage{amssymb}

\usepackage{float}

\begin{document}

\title{On the Capacity of Bandlimited Optical Intensity Channels with Gaussian Noise}

\author{Jing~Zhou
        and~Wenyi~Zhang,~\IEEEmembership{Senior~Member,~IEEE}
\thanks{This work was supported in part by the National Key Basic Research Program of China under Grant 2013CB329205, by the National Natural Science Foundation of China under Grant 61379003, and by the Fundamental Research Funds for the Central Universities under Grants WK2100060020 and WK3500000003.

Jing Zhou is with the Department
of Electronic Engineering and Information Science, University of Science and Technology of China, Hefei,
China, and with the Air Force Aviation University, Zibo, China (e-mail: jzee@ustc.edu.cn).

Wenyi Zhang is with the Key Laboratory of Wireless-Optical Communications, Chinese Academy of Sciences, and with the Department
of Electronic Engineering and Information Science, University of Science and Technology of China, Hefei,
China (e-mail: wenyizha@ustc.edu.cn).}}

\maketitle

\begin{abstract}

We determine lower and upper bounds on the capacity of bandlimited optical intensity channels (BLOIC) with white Gaussian noise. Three types of input power constraints are considered: 1) only an average power constraint, 2) only a peak power constraint, and 3) an average and a peak power constraint. Capacity lower bounds are derived by a two-step process including 1) for each type of constraint, designing admissible pulse amplitude modulated input waveform ensembles, and 2) lower bounding the maximum achievable information rates of the designed input ensembles. Capacity upper bounds are derived by exercising constraint relaxations and utilizing known results on discrete-time optical intensity channels. We obtain degrees-of-freedom-optimal (DOF-optimal) lower bounds which have the same pre-log factor as the upper bounds, thereby characterizing the high SNR capacity of BLOIC to within a finite gap. We further derive intersymbol-interference-free (ISI-free) signaling based lower bounds, which perform well for all practical SNR values. In particular, the ISI-free signaling based lower bounds outperform the DOF-optimal lower bound when the SNR is below 10 dB.
 \end{abstract}


\section{Introduction}

\subsection{Background and Related Work}

\IEEEPARstart{I}{ntensity} modulation and direct detection (IM/DD) is widely used in optical communications. In IM/DD, information is conveyed by the varying intensity of transmitted optical signal, and the receiver detects the intensity of the received signal. There have been extensive studies on design and realization of IM/DD based optical wireless communication systems (see \cite{H052,KU14} and references therein). This paper focuses on a simplified IM/DD channel model, known as the optical intensity channel with Gaussian noise. This model is suitable for some kinds of IM/DD systems, e.g., indoor free space optical communications \cite{H052,HK04}. In optical intensity channels, typically the average and/or peak optical power is constrained\footnote{Since the optical intensity is the optical power transferred per unit area, a constraint on optical power is also a constraint on intensity. In this paper we use `power' to refer to the optical power unless otherwise specified.} due to safety reasons and practical considerations. Therefore the optical intensity channel we considered has two fundamental differences with electrical channel models like the classical AWGN channel: 1) a nonnegativity constraint on the input signal, 2) a different input cost metric. It is clear that typical signaling methods for electrical channels cannot be applied directly in optical intensity channels because of these differences.

A number of information theoretic studies on optical intensity channels have been reported, mostly on discrete-time optical intensity channels (DTOIC); see, e.g., \cite{HK04,LMW09,FH09,FH10,CMA15,JWWD16}. For DTOIC with only an average power constraint, the exact capacity is still unknown, whereas tight upper and lower bounds have been established. When the input is further bounded from above due to a peak power constraint, the optimal input distribution and the capacity can be numerically computed \cite{Smith71,CHK05}. Moreover, recent works have provided systematic results on capacity-achieving input design for DTOIC \cite{FH09}, \cite{CMA15}. Only a few information theoretic studies have considered continuous-time optical intensity channel models. In \cite{YK02}, sphere packing based capacity upper bounds for multicarrier optical intensity channels (MCOIC) were established.
Capacity bounds for bandlimited optical intensity channels (BLOIC) with an average power constraint were studied in \cite{HK04}. In another aspect, the design of ISI-free signaling over BLOIC was studied in \cite {H07,TAKBIM12,CKA14}, while the information rate was not considered therein.

\subsection{Channel Model and Motivation of Our Study}

The BLOIC model considered in this paper is
\begin{equation}
\label{eqn:BLOIC}
Y(t)=X(t)+Z(t),\mspace{16mu}X(t)\ge 0
\end{equation}
where $X(t)$ is bandlimited to $\mathcal{W}\triangleq[-W,W]$, $Z(t)$ is white Gaussian noise (the definition follows that in \cite{Lapidoth}) of one-sided power spectral density $N_0$ with respect to $\mathcal{W}$. The bandwidth constraint of BLOIC is due to the optoelectronic components and multipath distortion \cite{H052,HK04}. In (\ref{eqn:BLOIC}), without loss of generality we set the channel gain (including, e.g., the responsivity of the photodiode and the optoelectronic conversion factor in IM/DD systems) to be unity, as in \cite{HK04}, \cite{LMW09}. The DTOIC
\begin{equation}
\label{eqn:DTOIC}
Y[n]=X[n]+Z[n],\mspace{16mu}X[n]\ge 0
\end{equation}
is the discrete-time analog of (\ref{eqn:BLOIC}), where $Z[n]\sim \mathrm{N}(0,\sigma^{2})$ ($\mathrm{N}(a,b)$ denotes Gaussian distribution with mean $a$ and variance $b$) is independent and identically distributed (i.i.d.).

To model the input constraint, we define the instantaneous powers as $\mathcal P(X(t))=x(t)$ and $\mathcal P(X[n])=x[n]$ for BLOIC and DTOIC, respectively. These definitions are different from those of the electrical power which is proportional to the squared amplitude of the signal. Then the average power is $\mathrm{E}\left[X[n]\right]$ in DTOIC. In BLOIC, the average power is defined as
\begin{equation}
\label{eqn:POWER}
\mathcal P_{\{X\}}=\lim_{T\rightarrow\infty}\frac{1}{2T} \mathrm E\left[\int_{-T}^T {X(t)\mathrm{d}t}\right].
\end{equation}
Here we follow the definition of [\ref{Lapidoth}, Definition 14.6.1]. 

For most classical channels such as the AWGN channel, the linear Gaussian channel or fading channels, we have the following relationship between the capacity of the continuous-time bandlimited channel model and its discrete-time analog with the same signal-to-noise ratio:
\begin{equation}
\label{eqn:BLDT}
\mathcal C_{\textrm{BL}}=\mathcal C_{\textrm{DT}}\cdot 2W\mspace{8mu}\textrm {transmissions per second}.
\end{equation}
This relationship can be established by orthogonal transforms, e.g., by Nyquist rate signaling/sampling or more rigorously by Karhunen-Lo\`{e}ve expansion and the 2WT theorem.\footnote{See \cite{G68}, \cite{LandauSlepianPollack62}. The 2WT theorem says that a channel with bandwidth $W$ has essentially $2WT$ degrees of freedom (DOFs) in a length-$T$ time interval, where $T\gg1/W$ and DOF is defined as the dimension of the signal space in that time interval.} Moreover, signaling schemes designed based on discrete-time models can be directly utilized in bandlimited communications by simple modulation/demodulation methods. Because of these facts, many studies based on classical channels reasonably consider only the discrete-time channel models, where the relationship between the discrete-time channel inputs (i.e. $x[n]$) and the continuous-time signal they represent (i.e. $x(t)$) is
\begin{equation}
\label{eqn:XidealBL1}
x(t)=\sum \limits _{n} x[n] \frac{\sin \pi (2Wt-n)}{\pi (2Wt-n)}.
\end{equation}

However, in this paper we emphasize that the aforementioned equivalent relationship does not hold between DTOIC and BLOIC. According to (\ref{eqn:XidealBL1}), it is easy to verify that the equivalent bandlimited waveform of a given nonnegative input sequence is not necessarily nonnegative everywhere. In other words, it is possible that an admissible input sequence in DTOIC corresponds to an inadmissible input waveform in BLOIC.\footnote{Note that the inadmissible waveforms can always be avoided in practice by proper engineering design, but at the expense of performance, as shown later.} Then it is clear that the capacity of BLOIC can not be obtained by solving the capacity problem of DTOIC and using (\ref{eqn:BLDT}). Therefore, when a bandwidth constraint exists, using DTOIC as the model of a continuous-time optical intensity channel is an \emph{oversimplification} of the capacity problem.

To the best of our knowledge, \cite{HK04} is the only information theoretic study directly on the BLOIC model (\ref{eqn:BLOIC}). This study is restricted to time-disjoint signaling (TDS) based on a finite (typically low) dimensional signal space model over a finite time interval. Using that model, the continuous-time channel is converted to a discrete-time vector Gaussian channel and the input nonnegativity constraint is correspondingly converted to an admissible region in the signal space. Capacity bounds are derived based on the converted vector Gaussian channel with the admissible input region. It is thus clear that the upper bounds obtained therein are only information rate upper bounds of specific TDS schemes, rather than capacity upper bounds for the BLOIC. Moreover, due to the finite time length of the signal space model, the bandwidth constraint can only be approximately satisfied by permitting an $\epsilon$-fractional out-of-band energy. Since the bandwidth is sensitive to $\epsilon$, varying $\epsilon$ causes the achievable spectral efficiency to vary significantly. When $\epsilon$ tends to zero the achievable spectral efficiency tends to zero. Even for a fixed $\epsilon$ and given bandwidth, the signaling rate is limited due to the poor time-frequency concentration of the rectangular basis function needed in the BLOIC signal space model. So the available DOFs of the bandlimited channel is hardly exploited in the most efficient way.

\subsection{Our Contribution}

In this paper, we study the fundamental limits of communication over the BLOIC under different types of input constraints. By designing admissible pulse amplitude modulated (PAM) input waveform ensembles with i.i.d. input symbols and lower bounding their maximum achievable information rates, we derive two kinds of capacity lower bounds: the DOF-optimal lower bounds and the ISI-free signaling based lower bounds. The DOF-optimal lower bounds achieve the optimal pre-log factor of the channel capacity by comparing with capacity upper bounds derived from constraint relaxation. Thus we characterize the high-SNR capacity of the BLOIC to within a finite gap. For example, the high-SNR asymptotic gap between the tightest lower and upper capacity bounds for the only average power constrained case is 4.34 dB in SNR. The ISI-free signaling is preferred in practical communication systems because of its low detection complexity. We show that the ISI-free signaling based lower bounds perform well for all practical SNR values, especially for low to moderate SNR. At high SNR, introducing a direct current (DC) bias in the signal design is shown to be very helpful for boosting the information rate. We also study the effects of different peak-to-average-power ratios and different types of modulation pulses (i.e. shaping filters) on the capacity lower bounds, and give several conjectures and discussions. All these results provide understanding on fundamental limits and signaling scheme design for bandlimited communications using IM/DD.

The remaining part of the paper is organized as follows. Section II presents our methods and main results. Section III gives comparisons among the capacity bounds based on numerical results. Section IV provides some discussions and two conjectures.

Throughout the paper the following notations are used: $p_X(x)$ denotes the probability density function (PDF) of $X$; $h(\cdot)$ stands for the differential entropy, i.e. $h(X)=-\int_{-\infty }^\infty {p_X(x)\log p_X(x)\mathrm{d}x}$; $I(\textsf Q,\textsf V)\triangleq I(X;Y)$ stands for the mutual information between input $X$ and output $Y$ of a channel with transition probability measure $\textsf V$ when $X$ has distribution $\textsf Q$; $\mathcal H[X(t)]$ stands for the differential entropy per DOF of the bandlimited waveform ensemble $X(t)$; $\mathcal I[X(t);Y(t)]$ stands for the mutual information per DOF between two bandlimited waveform ensembles $X(t)$ and $Y(t)$; $\mathcal C_\textrm{A}^\textrm{B}$ denotes the capacity of channel A under constraint B; $\mathcal R_\textrm{A}^\textrm{B}$ denotes the maximum achievable information rate of a constrained signaling scheme B over channel A. Boldface is used to denote matrices and vectors. Table I lists some abbreviations used in this paper.
\begin{table}[tbp]
\renewcommand\arraystretch{1.8}
\centering
\caption{Abbreviations}
\begin{tabular}{l|l}
\hline
AP &Average power\\\hline
BLAWGN &Bandlimited AWGN\\\hline
BLOIC &Bandlimited optical intensity channel\\\hline
DC &Direct current\\\hline
DOF &Degrees of freedom\\\hline
DTAWGN &Discrete-time AWGN\\\hline
EPI & Entropy power inequality\\\hline
IM/DD &Intensity modulation and direct detection\\\hline
i.i.d. &Independent and identically distributed\\\hline
ISI &Intersymbol interference\\\hline
MCOIC &Multicarrier optical intensity channel\\\hline
PAM &Pulse amplitude modulation\\\hline
PAPR &Peak-to-average-power-ratio\\\hline
PC &Power constraint\\\hline
PDF &Probability density function\\\hline
PL pulse &Parametric linear pulse\\\hline
PNR &Peak-to-noise ratio\\\hline
PP &Peak power\\\hline
PSWF & Prolate spheroidal wave function\\\hline
SC pulse & Spectral-cosine pulse\\\hline
SNR &Signal-to-noise ratio\\\hline
TDS &Time-disjoint signaling\\\hline
\end{tabular}
\end{table}

\section{Methods and Results}

\subsection{Preliminaries and Basic Methods}

The different input power constraints in the BLOIC considered in this paper are given in Table II. We use $r$ to denote the peak-to-average-power-ratio (PAPR), which is the ratio of the maximum allowed peak power (PP) to the maximum allowed average power (AP). We further use PC as a general notation for these power constraints when a general discussion on them is needed. We denote a BLOIC under AP constraint as AP-BLOIC, and so on. For the DTOIC and the BLAWGN channel similar notations are used.

\begin{table}[tbp]
\renewcommand\arraystretch{2}
\centering
\caption{Different Input Power Constraints in BLOIC}
\begin{tabular}{l|l}
\hline
\textbf{Power Constraint}& \textbf{Definition}\\ \hline
AP &$\mathcal P_{\{X\}}\leq \mathcal{E}, 0\leq x(t)\leq \infty $\\  \hline 
PP &$0 \leq x(t)\leq \mathcal{A}$ \\\hline
PAPR &$\mathcal P_{\{X\}}\leq \mathcal{E},\mspace{8mu} 0\leq x(t)\leq r\mathcal{E}$ \\\hline
\end{tabular}
\end{table}

In a bandlimited channel the input and output are random waveforms, while the input is drawn from a given waveform ensemble. Following Shannon \cite{Shannon48}, we define the entropy per DOF of an input ensemble $X(t)$ through the distribution of its Nyquist sample sequence as
\begin{equation}
\label{eqn:BLsampleentropy}
\mathcal H [X(t)]=\mathop {\lim }\limits_{N\to \infty }\frac{1}{2N+1} \int {p_{\rm{\bf X}}({\rm {\bf x}})\log \frac{1}{p_{\bf X}({\rm {\bf x}})}
\mathrm{d}\bf x}
\end{equation}
where ${\rm {\bf X}}=X_{-N}, \ldots, X_0, \ldots, X_{N} $ is the Nyquist sample sequence of $X(t)$. For example, for $Z(t)$ defined in (\ref{eqn:BLOIC}) we have $\mathcal H[Z(t)]=\log \sqrt {2\pi eN_0 W}$. The capacity of a bandlimited channel with input ensemble $X(t)$ and output $Y(t)$ is defined as
\begin{equation}
\label{eqn:CBC1}
\mathcal C=2W\cdot\mathop {\max }\limits_{p_{\bf X}({\bf x})} \mathcal I[X(t);Y(t)]
\end{equation}
where $\mathcal I[X(t);Y(t)]$ is the mutual information per DOF between $X(t)$ and $Y(t)$:
\begin{align}
\label{eqn:CBC}
&\mathcal I[X(t);Y(t)]\notag\\
&=\mathop {\lim}\limits_{N\to \infty } \frac{1}{2N+1}\int\!\!\!\int {p_{\bf {X,Y}}({\bf x},{\bf y})\log \frac{p_{\bf {X,Y}}({\bf x},{\bf y})}{p_{\bf {X}}({\bf x})p_{\bf {Y}}({\bf y})}} \mathrm{d}{\bf x}\mathrm{d}{\bf y},
\end{align}
where ${\rm {\bf X}}=X_{-N}, \ldots, X_0, \ldots, X_{N} $ and ${\rm {\bf Y}}=Y_{-N}, \ldots,Y_0, $ $\ldots,  Y_{N}$, which are the Nyquist samples of $X(t)$ and $Y(t)$, respectively. The maximum achievable information rate $\mathcal R$ of a specific signaling scheme has the same definition as $\mathcal C$ except that the input ensemble $X(t)$ must be generated using that signaling scheme.

Our achievability results (lower bounds) for the BLOIC are derived by two basic steps:
\begin{enumerate}
\renewcommand{\labelenumii}{(\arabic{enumii})}
\item Designing an admissible input waveform ensemble satisfying certain constraints.
\item Lower bounding the maximum achievable information rate of the designed input ensemble.
\end{enumerate}
In particular, we design PAM input ensembles with i.i.d. input symbols as
\begin{equation}
\label{eqn:PAM}
X_\textrm{PAM}(t)=\sum_{i} X_{i}g(t-iT_\textrm {s}), \mspace{16mu}X_\textrm{PAM}(t)\ge 0
\end{equation}
to accomplish the first step, where the modulation pulse $g(t)$ is a real $\mathcal L_2$ function (i.e. a finite-energy signal) bandlimited to $\mathcal W$. The design includes reasonable choices of the symbol rate $1/T_\textrm s$, the input symbol distribution $p_X(x)$, and the pulse $g(t)$.

\begin{figure}
\centering
\includegraphics[width=3.8in,height=2.85in]{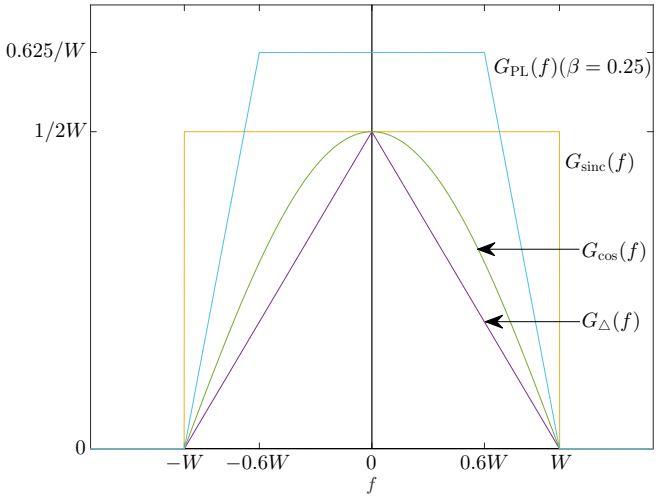}
\caption{The Fourier transforms of the pulses used.}
\label{Fignew1}
\end{figure}

Table III lists the pulses used in our results, including the sinc pulse, the S2 pulse \cite{TAKBIM12}, the spectral-cosine (SC) pulse, and the (first order) parametric linear (PL) pulse \cite{BD04} (the definition of parameters $\mathcal S_\textrm N$ and $\mathcal G$ will be given in (\ref{eqn:chi})--(\ref{eqn:SN})). Fig. \ref{Fignew1} shows the Fourier transform of these pulses. To simplify the proof of results we normalize the sinc, the S2, and the SC pulses to make them satisfy
\begin{equation}
\label{eqn:Norm}
\int _{-\infty} ^{\infty} g(t)\mathrm d t =\frac{1}{2W},
\end{equation}
and normalize the PL pulse to make it satisfy the following definition.
\begin{table*}[tbp]
\renewcommand\arraystretch{2.8}
\centering
\caption{List of the Pulses Used}
\begin{tabular}{l|l|l}
\hline
\textbf{Name}&\textbf{Notation and Definition} & \textbf{Remarks}\\ \hline
Sinc &$g_{\textrm {sinc}}(t)=\textrm {sinc}(2Wt)=\frac{\sin 2\pi Wt}{2\pi Wt}$&$\mathcal S_\textrm N =\infty$, $\mathcal G=1$\\  \hline
S2  &$g_\triangle(t)=\frac{1}{2}\left(\textrm{sinc}(Wt)\right)^2=\frac{\sin ^2\pi Wt}{2(\pi Wt)^2}$&$\mathcal S_\textrm N =1$, $\mathcal G=\frac{1}{e^2}$ \\\hline
SC &$g_\textrm {cos}(t)=\textrm {sinc}\left(2Wt-\frac{1}{2}\right)+\textrm {sinc}\left(2Wt+\frac{1}{2}\right)=\frac{2\cos2\pi Wt}{\pi\left( 1-16W^2t^2\right)}$&$\mathcal S_\textrm N =\frac{4}{\pi}$, $\mathcal G=\frac{1}{4}$ \\\hline
PL &$g_\textrm{PL}(t)=\textrm {sinc}\left(\frac{2Wt}{1+\beta}\right)\textrm {sinc}\left(\frac{2\beta Wt}{1+\beta}\right)=\frac{\sin\left(\frac{1}{1+\beta}2\pi Wt \right) \sin\left(\frac{\beta}{1+\beta}2\pi Wt\right)}{\beta\left( \frac{2\pi Wt}{1+\beta}\right)^2}, \mspace{6mu}\beta \in (0,1] $ & \\\hline
\end{tabular}
\end{table*}

\emph{Definition 1}: A normalized Nyquist pulse $g_\beta(t)$ with roll off factor $\beta$ is a real $\mathcal L_2$ function which is bandlimited to $\mathcal W$ and satisfies
\begin{equation}
\label{Nyquist}
g_\beta(nT_\mathrm 0)=\delta[n]
\end{equation}
where $T_\mathrm 0 =\frac{1+\beta}{2W}$, and $\delta[n]$ is the unit impulse.

\emph{Note}: Letting $G_\beta(f)$ be the Fourier transform of $g_\beta(t)$, it is easy to show that $G_\beta(f)$ satisfies
\begin{equation}
\label{norm}
G_\beta(0)=\int _{-\infty}^{\infty}g_\beta(t)\mathrm d t=T_\mathrm 0.
\end{equation}

Our lower bounds can be categorized as DOF-optimal lower bounds and ISI-free signaling based lower bounds, which are given in Sec. II-B and Sec. II-C, respectively. The basic idea for deriving DOF-optimal lower bounds is due to Shannon's derivation of capacity bounds for PP-BLAWGN channel in his 1948 landmark paper \cite{Shannon48}. The ISI-free signaling based lower bounds are derived by designing admissible ISI-free signaling schemes and considering the capacity of the equivalent discrete-time memoryless channel models.

We will present some general lower bounds which holds for all $g(t)$ or $p_X(x)$ satisfying certain constraints, and then use some specific $g(t)$ or $p_X(x)$ to get specific lower bounds. Some of the general lower bounds are given with respect to two parameters which are
\begin{equation}
\label{eqn:chi}
\mathcal G\triangleq\exp\left(\frac{1}{W}\int_{0}^{W} {\log\left|2W\cdot {G(f)} \right|^2 \mathrm{d}f}\right),
\end{equation}
\begin{equation}
\label{eqn:S}
\mathcal S(\tau)\triangleq\max\limits_{t\in[0,\tau]}\sum\limits_{i=-\infty}^{\infty}\left|g\left(t-i\tau\right)\right|,
\end{equation}
where $G(f)$ is the Fourier transform of $g(t)$. General lower bounds of this flavor were introduced for the PP-BLAWGN channel in \cite{Shamai88}, where Shannon's 1948 lower bound was tightened by optimizing the modulation pulse used. For brevity we further define
\begin{equation}
\label{eqn:SN}
\mathcal S_\mathrm N \triangleq \mathcal S\left(\frac{1}{2W}\right), \mspace{10mu} \mathcal S_\beta \triangleq \mathcal S\left(\frac{1+\beta}{2W}\right).
\end{equation}

The converse results are given in Sec. II-D, where the bounding technique used is also based upon \cite{Shannon48}.

\subsection{DOF-Optimal Capacity Lower Bounds}

The following lemma from \cite{Shannon48} plays a crucial role in deriving the results in this subsection.

\emph{Lemma 1}: If an ensemble of waveform $X_\textrm{I}(t)$ bandlimited to $\mathcal{W}$ is filtered by $G(f)$, then the entropy per DOF of the output ensemble is
\begin{equation}
\label{eqn:entropylossLF}
\mathcal H[X_\textrm{O}(t)]=\mathcal H[X_\textrm{I}(t)]+\frac{1}{2W}\int_{0}^{W} {\log \left| {G(f)} \right|^2 \mathrm{d}f}.
\end{equation}

\begin{IEEEproof}See [\ref{Middleton}, Chapter 6.4].
\end{IEEEproof}

All the results in this subsection are derived using i.i.d. Nyquist rate input ensembles as
\begin{equation}
\label{eqn:PAM-Nyquist}
X_\textrm{PAM}(t)=\sum_{i} X_{i}g\left(t-\frac{i}{2W}\right), \mspace{16mu}X_\textrm{PAM}(t)\ge 0.
\end{equation}
A symbol rate no less than the Nyquist rate is necessary to exploit all the available DOFs of bandlimited channels in the high SNR regime \cite{FU98}.

\emph{Lemma 2}: The maximum achievable information rate achieved by the i.i.d. Nyquist rate ensemble $X_\textrm{PAM}(t)$ in (\ref{eqn:PAM-Nyquist}) transmitted over the BLOIC can be lower bounded by
\begin{equation}
\label{eqn:LB-PAM}
\mathcal R_{\textrm{PAM}} \ge W\log \left(1+\frac{\mathcal G \exp{(2h(X))}}{2 \pi e N_0 W}\right)
\end{equation}
where $\mathcal G$ is defined as (\ref{eqn:chi}).

\begin{IEEEproof}
Consider an ergodic or cyclostationary ensemble of waveform $X(t)$ bandlimited to $\mathcal{W}$. For an additive noise channel bandlimited to $\mathcal W$ the information rate

\begin{align}
\label{eqn:LBCANC}
\mathcal R
&=2W\cdot\mathcal I\left[X(t);Y(t)\right]\notag\\
&=2W\cdot\left(\mathcal H [Y(t)]-\mathcal H \left[Y(t)|X(t)\right]\right)\notag\\
&=2W\cdot\left(\mathcal H\left[X(t)+Z(t)\right]-\mathcal H[Z(t)]\right)
\end{align}
is achievable by using the ensemble $X(t)$.\footnote{Note that information measures of waveforms, $\mathcal H$ and $\mathcal I$, have the same properties as the differential entropy $h(\cdot)$ and mutual information $I(X;Y)$ of scalar variables, respectively, if the limits in (\ref{eqn:BLsampleentropy}) and (\ref{eqn:CBC}) exist.} Using the vector version of the entropy power inequality (EPI) \cite{CoverThomas06}
\begin{equation}
\label{eqn:EPI}
e^{\frac2N h(\bf X+\bf Y)}\ge e^{\frac2N h(\bf X)}+e^{\frac2N h(\bf Y)},
\end{equation}
the information rate given by (\ref{eqn:LBCANC}) can be lower bounded as
\begin{align}
\label{eqn:EPILB-ANC}
\mathcal R
&\ge W\log \left(1+e^{2\mathcal H[X(t)]-2\mathcal H[Z(t)]}\right)\notag\\
&=W\log \left(1+\frac{\exp\left(2\mathcal H[X(t)]\right)}{2\pi e N_\mathrm 0 W}\right).
\end{align}
Since $X_i$ is i.i.d., the ensemble $X_\textrm{PAM}(t)$ is cyclostationary and the information rate (\ref{eqn:EPILB-ANC}) (replacing $X(t)$ with $ X_\textrm{PAM}(t)$) is achievable. Now we evaluate $\mathcal H [X_\textrm{PAM}(t)]$. We note that the ensemble $X_\textrm{PAM}(t)$ as (\ref{eqn:PAM-Nyquist}) can be obtained by filtering an ideal bandlimited ergodic ensemble $X_{\textrm{sinc}}(t)=\sum _{i} X_i g_{\textrm {sinc}}(t-i/2W)$ by $2W\cdot G(f)$. This is because if $g_{\textrm{sinc}}(t)$ is filtered by $2W\cdot G(f)$, the output is $g(t)$ ($G_\textrm{sinc}(f)\cdot 2W \cdot G(f)=G(f)$ as $G_{\textrm {sinc}}(f)$ equals $1/2W$ within $\mathcal W$). Then by using Lemma 1 and noting that $\mathcal H[X_{\textrm{sinc}}(t)]=h(X)$ (we omit the index of $\{X_i\}$ since they are i.i.d.), we have $\mathcal H [X_\textrm{PAM}(t)]=h(X)+\frac{1}{2}\log \mathcal G$. Combining this with (\ref{eqn:EPILB-ANC}) completes the proof.
\end{IEEEproof}

\emph{Theorem 1}: The capacity of the AP-BLOIC is lower bounded by
\begin{equation}
\label{eqn:GLB-AP-BLOIC}
\mathcal C_{\textrm{BLOIC}}^{\textrm {AP}} \ge W\log \left(1+\frac{\mathrm {exp}\left(2h(X) \right)}{2\pi e^3 N_0 W}\right)
\end{equation}
for any $p_X(x)$ satisfying $\mathrm{E}[X]=\mathcal{E}$ and $p_X(x)=0$ for $x<0$.

\begin{IEEEproof}The proof is given in Appendix A.
\end{IEEEproof}

Theorem 1 is obtained by employing the pulse $g_\triangle(t)$ and nonnegative input symbols $X_i$ in (\ref{eqn:PAM-Nyquist}) and lower bounding the information rate achieved.
The maxentropic distribution of a nonnegative random variable with a given expected value $\mathcal E$ is the exponential distribution \cite{CoverThomas06}:
\begin{equation}
\label{eqn:EPDF}
p_X(x)={\cal E}^{-1}e^{-{x} \mathord{\left/ {\vphantom {{x} {\cal
E}}} \right. \kern-\nulldelimiterspace} {\cal E}},
\end{equation}
whose differential entropy is
\begin{equation}
\label{eqn:hE}
h(X)=\log e {\cal E}.
\end{equation}
Substituting (\ref{eqn:hE}) into (\ref{eqn:GLB-AP-BLOIC}) yields the following corollary.

\emph{Corollary 1}:
\begin{equation}
\label{eqn:LB-AP-BLOIC}
\mathcal C_{\textrm{BLOIC}}^{\textrm {AP}} \ge W\log \left(1+\frac{1}{2\pi e}\frac{{\mathcal E}^2}{N_0 W}\right).
\end{equation}

We call (\ref{eqn:LB-AP-BLOIC}) the Exp-S2 lower bound since it is based on an input ensemble using exponential symbol distribution and S2 pulse. While (\ref{eqn:LB-AP-BLOIC}) is the tightest bound we could obtain from Theorem 1, the general bound (\ref{eqn:GLB-AP-BLOIC}) still has its own merit since it can be used to evaluate the performance of more practical input symbol distributions (e.g., a uniform distribution).

\emph{Theorem 2}: The capacity of the PP-BLOIC is lower bounded by
\begin{equation}
\label{eqn:LB-Shamai-General-PP-BLOIC}
\mathcal C_{\textrm{BLOIC}}^{\textrm{PP}}\ge W\log \left(1+\frac{\mathcal G}{2\pi e \mathcal S_\mathrm N^2} \frac {\mathcal {A}^2}{N_{0}W}\right)
\end{equation}
where $\mathcal G$ and $\mathcal S_\mathrm N$ are defined as (\ref{eqn:chi}) and (\ref{eqn:SN}), respectively, with respect to an arbitrary modulation pulse $g(t)$ satisfying (\ref{eqn:Norm}).

\begin{IEEEproof}The proof is given in Appendix B.
\end{IEEEproof}

Theorem 2 is derived using an uniform input symbol distribution which is the maxentropic distribution for bounded random variables without further constraint.
If we employ $g_\textrm{cos}(t)$ as the modulation pulse and note that for $g_{\textrm {cos}}(t)$ we have $\mathcal S_\textrm N=4/\pi$ and $\mathcal G=1/4$ (see \cite{Shamai88}), we get the following corollary called the Unif-cos lower bound which is a suboptimal example of (\ref{eqn:LB-Shamai-General-PP-BLOIC}).

\emph{Corollary 2}:
\begin{equation}
\label{eqn:LB-Shamai-PP-BLOIC}
\mathcal C_{\textrm{BLOIC}}^{\textrm{PP}}\ge W\log \left(1+\frac{\pi}{128 e} \frac {\mathcal {A}^2}{N_{0}W}\right).
\end{equation}

Theorem 2 and Corollary 2 can be viewed as parallel results of \cite{Shamai88} (in which $g_\textrm{cos}(t)$ is proposed) on lowpass PP-BLAWGN channel.

\emph{Theorem 3}: The capacity of the PAPR-BLOIC is lower bounded by
\begin{equation}
\label{LB-APPP-General}
\begin{split}
\mathcal C_{\textrm{BLOIC}}^{\textrm{PAPR}}(r) \ge \mspace{382mu}\\
\begin{cases}
W\log \bigg( 1+ \frac { \mathcal G r^2 \exp\left(\frac{2\mathcal S_\mathrm N-r\mathcal S_\mathrm N+r}{r}\mu\right)}{2\pi e \mathcal S_\mathrm N^2 }\left( \frac{1-e^{-\mu}}{\mu}\right)^2 \frac {\mathcal E^2}{ N_0W}\bigg),r>2\\
W\log \left(1+\frac{\mathcal G r^2}{2\pi e \mathcal S_\mathrm N^2} \frac {\mathcal {E}^2}{N_{0}W}\right),\mspace{160mu}0<r\leq2,
\end{cases}
\end{split}
\end{equation}
where $r$ is the PAPR, $\mathcal G$ and $\mathcal S_\mathrm N$ are defined as (\ref{eqn:chi}) and (\ref{eqn:SN}), respectively, with respect to an arbitrary modulation pulse $g(t)$ satisfying (\ref{eqn:Norm}), and $\mu$ is the unique solution to
\begin{equation}
\frac{2\mathcal S_\mathrm N-r\mathcal S_\mathrm N+ r}{2 r}=\frac {1}{\mu}-\frac {e^{-\mu}}{1-e^{-\mu}},
\end{equation}

\begin{IEEEproof}The proof is given in Appendix C.
\end{IEEEproof}

In the proof of Theorem 3, we let the input symbol distribution be a truncated exponential distribution which is the maxentropic distribution of a nonnegative random variable with a given expected value and an upper bound. This distribution was used for bounding the capacity of the DTIOC in \cite{LMW09}. See Appendix C for details.

\emph{Note:} For the PAPR-DTOIC, when $r\leq2$ the AP constraint becomes inactive and the capacity is equal to that of the PP-DTOIC with the same PP constraint \cite{LMW09}. For the PAPR-BLOIC, however, it is nontrivial to find out the PAPR transition point at which the AP constraint becomes inactive. Note that $r=2$ is only the transition point in (\ref{LB-APPP-General}), which is not the capacity.

By employing the pulse $g_\triangle (t)$ in (\ref{LB-APPP-General}) we get the following specific lower bound, called the TE-S2 (truncated-exponential-S2) lower bound, which is a suboptimal example of (\ref{LB-APPP-General}).

\emph{Corollary 3}:
\begin{equation}
\begin{split}
\label{eqn:LB-CPAPR-BLOIC}
\mathcal C_{\textrm{BLOIC}}^{\textrm{PAPR}}(r)\ge\mspace{360mu}\\
\begin{cases}
W\log \left( 1+ \frac {r^2 e^{2\mu/r}}{2\pi e^3 }\left( \frac{1-e^{-\mu}}{\mu}\right)^2 \frac {\mathcal {E}^2}{N_0W}\right),\mspace{8mu} &r>2\\
W\log \left(1+\frac{r^2}{2\pi e^3} \frac {\mathcal {E}^2}{N_{0}W}\right),&0<r\leq2,
\end{cases}
\end{split}
\end{equation}
where $r> 2$, and $\mu$ is the unique solution to
\begin{equation}
\label{eqn:alpha}
\frac{1}{r}=\frac {1}{\mu}-\frac {e^{-\mu}}{1-e^{-\mu}}.
\end{equation}

\begin{IEEEproof}The proof is given in Appendix C.
\end{IEEEproof}

\subsection{ISI-Free Signaling based Capacity Lower Bounds}

The results in this subsection are given in the form of lower bounds on $\mathcal R_\textrm {BLOIC}^\textrm{PC, IFS}$, which is the maximum achievable information rate of ISI-free signaling over the PC-BLOIC. Of course, they are also lower bounds on the capacity of the PC-BLOIC.

ISI-free signaling avoids ISI by using modulation pulses that satisfy the Nyquist criterion \cite{Lapidoth}. It may use Nyquist pulses (e.g., raised-cosine pulse) and a direct-sampling detector, or alternatively use the so-called $T$-orthogonal pulses (e.g., root raised-cosine pulse) and a matched filter detector. ISI-free signaling achieves the Nyquist rate only when the sinc pulse is used.

For the BLOIC, \cite{H07} recognizes two important facts on ISI-free signaling using \textit{nonnegative} pulses:
\begin{enumerate}
\renewcommand{\labelenumii}{(\arabic{enumii})}
    \item ISI-free signaling is impossible when a matched filter receiver is used.
    \item ISI-free signaling is possible when a direct-sampling receiver is used. The maximum symbol rate is a half of the Nyquist rate, achieved by employing the pulse $g_\triangle (t)$.
\end{enumerate}
The derivations of all the bounds in the following two theorems use the second fact, i.e., employing i.i.d. PAM signaling as (\ref{eqn:PAM}) with modulation pulse $g_\triangle(t)$ and letting $T_\textrm s=T_\textrm 0=\frac{1}{W}$.

\emph{Theorem 4}: The maximum achievable information rates of ISI-free signaling over the AP-BLOIC, the PP-BLOIC, and the PAPR-BLOIC, are lower bounded by (\ref{eqn:LB-ISI}), (\ref{eqn:LB-ISI-U-1}), and (\ref{eqn:LB-ISI-U}), respectively:
\begin{align}
\label{eqn:LB-ISI}
\mathcal R_{\textrm{BLOIC}}^{\textrm {AP},\mspace{4mu}\textrm {IFS}} &\ge \frac{W}{2}\log \left(1+\frac{e}{2\pi}\frac{{\mathcal E}^2}{N_0 W}\right);\\
\label{eqn:LB-ISI-U-1}
\mathcal R_{\textrm{BLOIC}}^{\textrm {PP},\mspace{4mu}\textrm {IFS}} &\ge \frac{W}{2}\log \left(1+\frac{1}{2\pi e}\frac{{\mathcal A}^2}{N_0 W}\right);\\
\label{eqn:LB-ISI-U}
\mathcal R_{\textrm{BLOIC}}^{\textrm {PAPR}, \mspace{4mu}\textrm {IFS}} &\ge
\begin{cases}
\frac{W}{2}\log \left( 1+ \frac {r^2 e^{2\mu/r}}{2\pi e }\left( \frac{1-e^{-\mu}}{\mu}\right)^2 \frac {\mathcal {E}^2}{N_0W}\right),\mspace{4mu}  r>2\\
\frac{W}{2}\log \left( 1+ \frac {r^2}{2\pi e } \frac {\mathcal {E}^2}{N_0W}\right), \mspace{84mu} 0<r\leq2,
\end{cases}
\end{align}
where $r$ is the PAPR, $\mu$ is the unique solution to (\ref{eqn:alpha}).

\begin{IEEEproof}The proof is given in Appendix D.
\end{IEEEproof}

Similar to the DOF-optimal bounds, the derivation of the bounds in Theorem 4 uses the maxentropic input symbol distributions for each type of constraint. We call (\ref{eqn:LB-ISI}), (\ref{eqn:LB-ISI-U-1}), and (\ref{eqn:LB-ISI-U}) the Exp-S2-IFS lower bound, the Unif-S2-IFS lower bound, and the TE-S2-IFS lower bound, respectively.  The following result (called the Geom-S2-IFS lower bound), however, uses a geometry distribution which has been proposed in \cite{FH10} for bounding the capacity of the DTOIC.

\emph{Theorem 5}:
\begin{equation}
\label{eqn:Gbound}
\mathcal R_{\textrm{BLOIC}}^{\textrm{AP},\mspace{4mu}\textrm {IFS}}\ge W \cdot \max\limits_l I(\textsf Q_\textrm{g}(l), \textsf V)
\end{equation}
where $\textsf Q_\textrm{g}(l)$ is a geometric distribution with PDF
\begin{equation}
\label{Geo}
p_X(x,l)=\sum \limits _{i=0}^{\infty} \frac{l}{l+\mathcal{E}}\left(\frac{\mathcal{E}}{l+\mathcal{E}}\right)^i \delta(x-il), \mspace{10mu} l>0,
\end{equation}
and $\textsf V$ is the transition probability of the channel $Y=X+Z$, where $Z\sim\textrm{N}(0,N_0W)$.

\begin{IEEEproof} The proof is given in Appendix E.
\end{IEEEproof}

The following result is based on DC-aided ISI-free signaling over the BLOIC \cite{TAKBIM12} whose symbol rate can surpass a half of, and even approach, the Nyquist rate.

\emph{Theorem 6}: The maximum achievable information rates of ISI-free signaling over the AP-BLOIC, PP-BLOIC, and PAPR-BLOIC are lower bounded by (\ref{LB-ISI-bias}) (\ref{LB-ISI-bias-1}), and (\ref{LB-ISI-bias-2}), respectively:
\begin{align}
\label{LB-ISI-bias}
\mathcal R_\textrm{BLOIC}^\textrm{AP, IFS}&\ge\sup \limits_{\beta \in (0,1]}\frac{W}{1+\beta}\log \left(1+\frac{2}{\mathcal S_\beta^2\pi e} \frac{\mathcal E^2}{N_0W} \right);\\
\label{LB-ISI-bias-1}
\mathcal R_\textrm{BLOIC}^\textrm{PP, IFS}&\ge\sup \limits_{\beta \in (0,1]}\frac{W}{1+\beta}\log \left(1+\frac{1}{2\mathcal S_\beta^2\pi e} \frac{\mathcal A^2}{N_0W} \right);\\
\label{LB-ISI-bias-2}
\mathcal R_\textrm{BLOIC}^\textrm{PAPR, IFS}&\ge
\begin{cases}
\sup \limits_{\beta \in (0,1]}\frac{W}{1+\beta}\log \bigg(1+\\
\mspace{20mu} \frac{r^2\exp\left(\frac{2\mathcal S_\beta-r\mathcal S_\beta+r}{r}\mu\right)}{2\mathcal S_\beta^2\pi e}
\left(\frac{1-e^{-\mu}}{\mu}\right)\frac{\mathcal E^2}{N_0W} \bigg),\mspace{2mu} r>2\\
\sup \limits_{\beta \in (0,1]}\frac{W}{1+\beta}\log \left(1+\frac{r^2}{2\mathcal S_\beta^2\pi e}\frac{\mathcal E^2}{N_0W} \right), \mspace{6mu} 0<r\leq2,
\end{cases}
\end{align}
where the parameter $\mathcal S_\beta$, defined as (\ref{eqn:SN}), is determined by the normalized Nyquist pulse $g_\beta(t)$ used, and $\mu$ is the unique solution to
\begin{equation}
\frac{2\mathcal S_\beta-r\mathcal S_\beta+ r}{2 r}=\frac {1}{\mu}-\frac {e^{-\mu}}{1-e^{-\mu}}.
\end{equation}

\begin{IEEEproof}The proof is given in Appendix F.
\end{IEEEproof}

By using PL pulse as $g_\beta(t)$ in Theorem 6, numerical lower bounds are given in Sec. III. The bounds obtained from (\ref{LB-ISI-bias}) and (\ref{LB-ISI-bias-1}) are called the Unif-PL-IFS lower bounds and the bound obtained from (\ref{LB-ISI-bias-2}) is called the TE-PL-IFS lower bound. The reason for choosing the PL pulse is as follows. In bias-aided ISI-free signaling over the BLOIC, there is a tradeoff between the required DC bias and the achieved symbol rate: a higher symbol rate requires a larger DC bias (and a larger power cost). This leads to a tradeoff between power and DOF, see Fig. \ref{fig:DC} in Sec. III. When the rate of ISI-free signaling is close to the Nyquist rate, the required DC bias increases sharply (cf. Fig. 4 of \cite{TAKBIM12}). Achieving ISI-free signaling at exactly the Nyquist rate is impossible because it requires an infinitely large DC bias. A general analysis on the optimal DC bias-symbol rate tradeoff of arbitrary Nyquist pulses is difficult. But for certain kind of parametric pulses (e.g., raised cosine pulse with a roll off factor $\beta$) this tradeoff has been numerically characterized in \cite{TAKBIM12} in which the PL pulse was shown to be a good choice in a variety of pulses.

\subsection{Capacity Upper Bounds}

The following lemma holds for all the input power constraints given in Table II.

\emph{Lemma 3}: The capacity of the PC-BLOIC with bandwidth $W$ is upper bounded by
\begin{equation}
\label{DT2BL}
\mathcal C_{\textrm {BLOIC}}^{\textrm {PC}}\leq \mathcal {C}_{\textrm {DTOIC}}^{\textrm {PC}, \sigma^2=N_0W}\cdot 2W \mspace{10mu} \textrm {transmissions per second}
\end{equation}
where $\mathcal {C}_{\textrm {DTOIC}}^{\textrm {PC}}$ is the capacity of the DTOIC under the same type of constraint with equal parameters as the PC-BLOIC.

\begin{IEEEproof} The proof is given in Appendix G.
\end{IEEEproof}

Combining Lemma 3 and known capacity upper bounds for the AP-DTOIC, the following capacity upper bounds for the AP-BLOIC are obtained.

\emph{Theorem 7}: The capacity of the AP-BLOIC is upper bounded by the following two bounds simultaneously:
\begin{equation}
\label{eqn:UB1-AP-BLOIC}
\mathcal C_{\textrm {BLOIC}}^{\textrm {AP}} \le W\log \left(\frac{e}{2\pi}\left( {\frac{{\cal E}}{\sqrt{N_{0}W} }+2}\right)^2 \right),
\end{equation}
\begin{equation}
\label{eqn:UB2-AP-BLOIC}
\begin{split}
&\mathcal C_{\textrm {BLOIC}}^{\textrm {AP}} \le\\
&\mathop {\sup }\limits_{\alpha \in [0,1]} W\left(\log
{\left( \frac{e}{2\pi}{\frac{{\cal E}^2}{N_{0}W}}
\right)^\alpha}-\log{(1-\alpha)^{2-2\alpha}\alpha^{3\alpha}} \right).
\end{split}
\end{equation}

\begin{IEEEproof}By Lemma 3 and the upper bounds for the AP-DTOIC from [\ref{bHK04}, eqn. (21)] (implicitly given therein) and [\ref{bCMA15}, eqn. (1)], (\ref{eqn:UB1-AP-BLOIC}) and (\ref{eqn:UB2-AP-BLOIC}) are obtained, respectively.
\end{IEEEproof}

\begin{table}[tbp]
\renewcommand\arraystretch{1.8}
\centering
\caption{Upper Bounds and Related DTOIC Results}
\begin{tabular}{l|l}
\hline
\textbf{Upper bounds }& \textbf{Related DTOIC result}\\\hline
SP UB3, Fig. \ref{fig:BLOIC-basic} &{[\ref{bFH10}, (11)]}\\\hline
Dual UB, Fig. \ref{fig:BLOIC-basic} &[\ref{bLMW09}, (28)]\\\hline
SP UB1 and SP UB2, Fig. \ref{fig:PP} &[\ref{bCMA15}, Theorem 1]\\\hline
Dual UB1, Fig. \ref{fig:PP} &[\ref{bLMW09}, (19)]\\\hline
Dual UB2, Fig. \ref{fig:PP} &[\ref{bLMW09}, (20)]\\\hline
Dual UB1, Fig. \ref{fig:PAPR} &[\ref{bLMW09}, (11)]\\\hline
Dual UB2, Fig. \ref{fig:PAPR} &[\ref{bLMW09}, (12)]\\\hline
\end{tabular}
\end{table}

\begin{figure*}
\centering
\includegraphics[width=5.6in,height=4.2in]{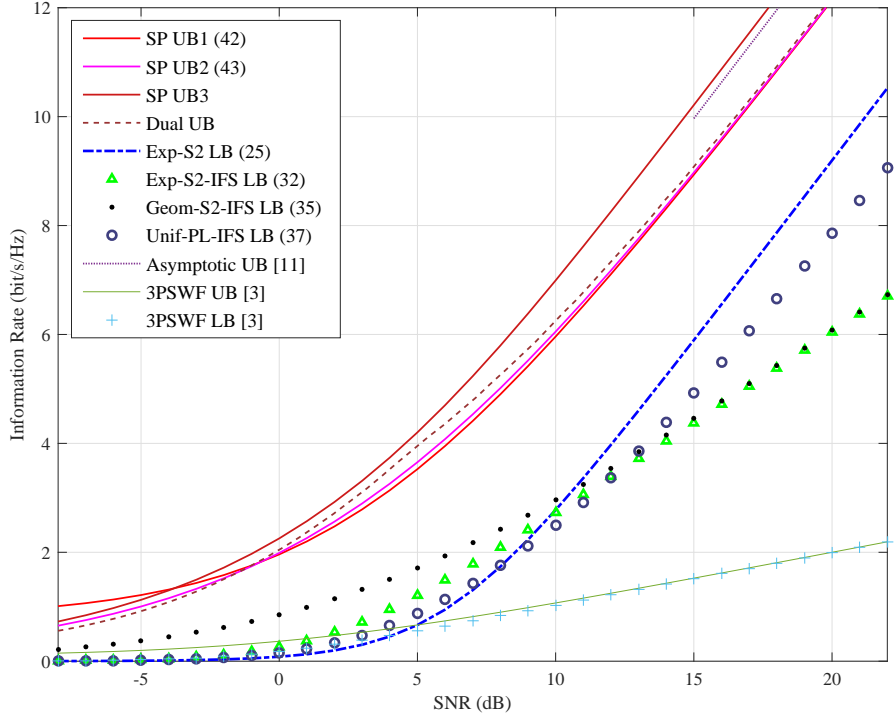}
\caption{Upper bounds (UB) and lower bounds (LB) for the capacity of the AP-BLOIC.}
\label{fig:BLOIC-basic}
\end{figure*}

\emph{Remark 1}: More capacity upper bounds for the AP-BLOIC can be obtained by other capacity upper bounds for the AP-DTOIC. Parallel results of Theorem 7 for the PP- and PAPR-BLOIC can be similarly obtained. Since the mathematical expressions of these results can be written out directly based on the corresponding DTOIC results, we only give them numerically in Sec. III, and list them in Table IV.
According to the type of the related DTOIC results, we categorize these upper bounds as sphere-packing based ones (SP UB) and duality based ones (Dual UB).

\section{Comparisons of Bounds}

In this section we give numerical evaluation of our results. In all figures the SNR and PNR of the BLOIC is defined as $\textrm{SNR}=\frac{\mathcal{E}}{\sqrt{N_0W}}$ and $\textrm{PNR}=\frac{\mathcal{A}}{\sqrt{N_0W}}$, respectively.

In Fig. \ref{fig:BLOIC-basic} our main results on the AP-BLOIC are plotted. At high SNR, it is shown that the Exp-S2 lower bound (\ref{eqn:LB-AP-BLOIC}) and the upper bound (\ref{eqn:UB1-AP-BLOIC}) are the tightest lower and upper bounds, respectively. Moreover, they have the same asymptotic slope and the high SNR asymptotic gap between them is 4.34 dB in SNR or 2.89 bit/s/Hz in spectral efficiency.

\begin{figure}
\centering
\includegraphics[width=3.8in,height=2.85in]{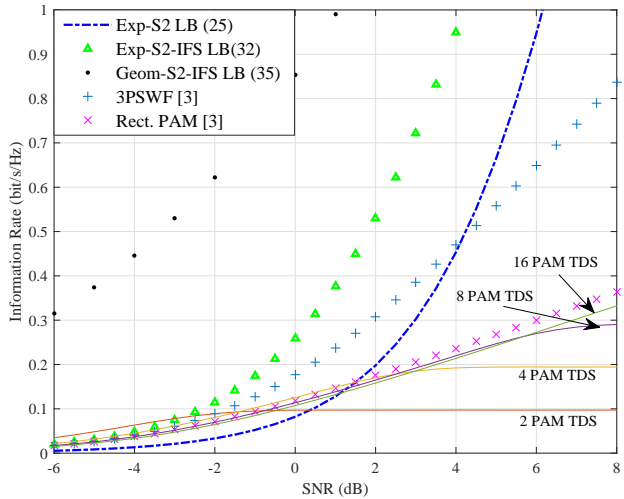}
\caption{Lower bounds on $\mathcal C_{\textrm {BLOIC}}^{\textrm {AP}}$: moderate SNR region. }
\label{fig:BLOIC-moderate}
\end{figure}

\begin{figure}
\centering
\includegraphics[width=3.8in,height=2.85in]{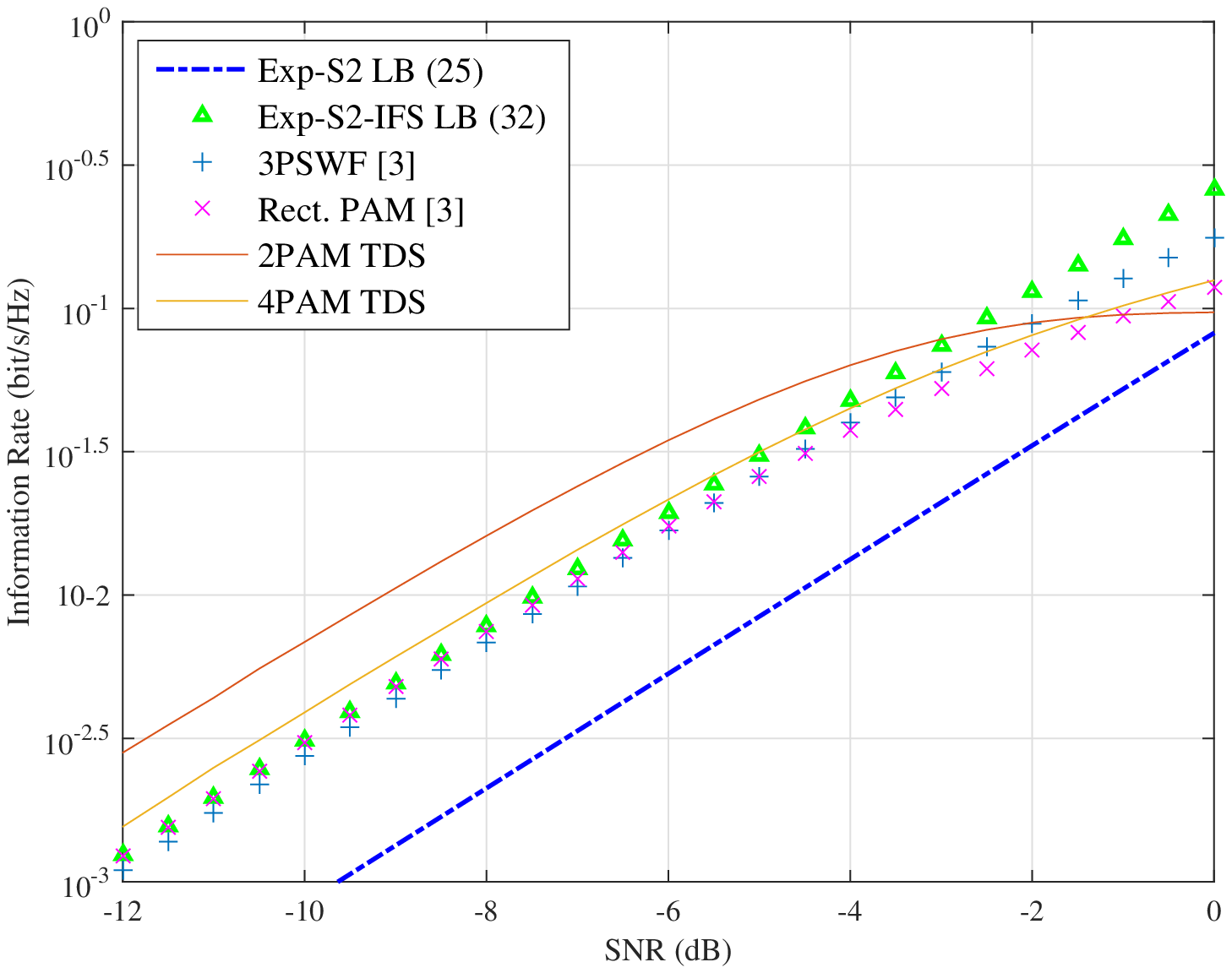}
\caption{Lower bounds on $\mathcal C_{\textrm {BLOIC}}^{\textrm {AP}}$: low SNR region.}
\label{fig:BLOIC-low}
\end{figure}

For comparison, the information rate bounds in \cite{HK04} for a specific TDS scheme called 3-PSWF, whose lower bound is the best among all examples in \cite{HK04}, are also plotted in Fig. \ref{fig:BLOIC-basic}. It is clear that the lower bound increases very slowly with SNR and the upper bound is not a capacity upper bound for the AP-BLOIC.\footnote{A $0.99$-fractional bandwidth definition is used in the evaluation of the performance of 3-PSWF (also the TDS based PAM results in Fig. \ref{fig:BLOIC-moderate} and Fig. \ref{fig:BLOIC-low}). Although the PSWF who achieve the best time-frequency concentration are used, the total DOF efficiency is dominated by the rectangular basis function which always exists in TDS signal space models.} Moreover, a high SNR asymptotic upper bound for the AP-BLOIC based on the result of \cite{YK02} is also shown and the details about this bound are given in Sec. IV.

An important observation is that ISI-free signaling performs well for all practical SNR values (e.g. in Fig. \ref{fig:BLOIC-basic} we show the SNR range $[-8, 22]$ in dB). At low to moderate SNR, all the ISI-free signaling based lower bounds outperform the Exp-S2 lower bound (\ref{eqn:LB-AP-BLOIC}), and the Geom-S2-IFS lower bound (\ref{eqn:Gbound}) is the tightest one. At high SNR, the Unif-PL-IFS lower bound (\ref{LB-ISI-bias}) obtained by DC bias aided ISI-free signaling achieves information rates close to the best known capacity lower bound obtained without ISI-free constraint.

Fig. \ref{fig:BLOIC-moderate} and Fig. \ref{fig:BLOIC-low} show the AP-BLOIC capacity lower bounds in the low to moderate SNR regime. The information rates of regular PAM constellations based (TDS) schemes of \cite{HK04} are also given for comparison. Only when the SNR is below 0 dB, the TDS schemes may have similar information rates compared with some of our lower bounds. The lower bound (\ref{eqn:Gbound}) stands out from all the results at low to moderate SNR. 

\begin{figure}
\centering
\includegraphics[width=3.8in,height=2.85in]{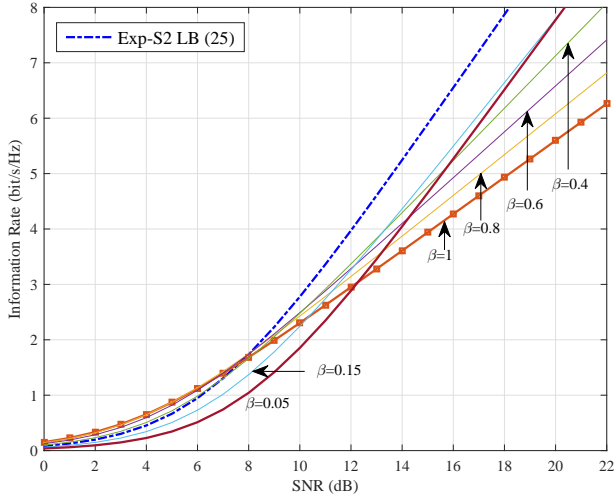}
\caption{Information rates of DC bias aided ISI-free signaling using $\mspace{30mu}$ $g_\textrm{PL}(t)$ under different roll-off factors.}
\label{fig:DC}
\end{figure}

Fig. \ref{fig:DC} shows the information rates of DC bias aided ISI-free signaling using $g_\textrm{PL}(t)$ under different roll-off factors (i.e. the RHS of (\ref{LB-ISI-bias}) excluding the supremum operation), and a tradeoff between low-SNR and high-SNR information rates for a given $\beta$ is clear. As practical systems always use a fixed $\beta$, a carefully chosen $\beta$ (typically from 0.15 to 0.4) may balance the performance for most practical SNR values.

\begin{figure}
\centering
\includegraphics[width=3.8in,height=2.85in]{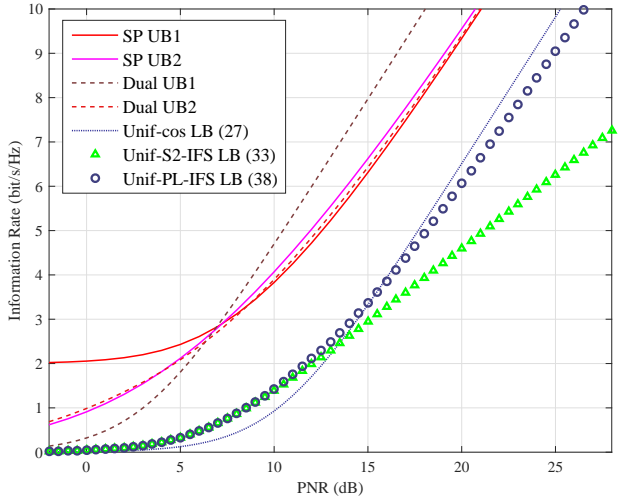}
\caption{Upper and lower bounds on the capacity of the PP-BLOIC.}
\label{fig:PP}
\end{figure}

\begin{figure}
\centering
\includegraphics[width=3.8in,height=2.85in]{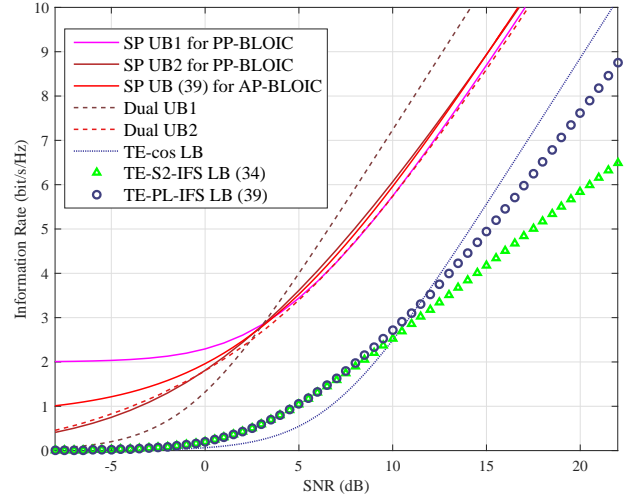}
\caption{Upper and lower bounds on the capacity of the PAPR-BLOIC, $r=2.5$.}
\label{fig:PAPR}
\end{figure}

Fig. \ref{fig:PP} and Fig. \ref{fig:PAPR} show the capacity bounds for the PP- and PAPR-BLOIC (where $r=2.5$), respectively. The behavior of these bounds are similar to that in the AP-BLOIC case. Note that at low SNR (\ref{eqn:LB-ISI-U-1}) and (\ref{eqn:LB-ISI-U}) are equal to (\ref{LB-ISI-bias-1}) and (\ref{LB-ISI-bias-2}), respectively.

\begin{figure}
\centering
\includegraphics[width=3.8in,height=2.85in]{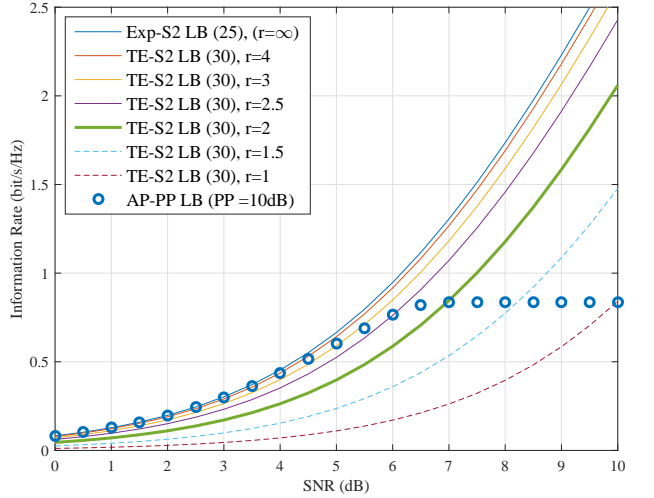}
\caption{Lower bounds on the capacity of the PAPR-BLOIC and the AP-PP-BLOIC.}
\label{fig:BLOIC-Peak}
\end{figure}

In Fig. \ref{fig:BLOIC-Peak} we show the lower bounds on the capacity of the PAPR-BLOIC given by (\ref{eqn:LB-CPAPR-BLOIC}) for different PAPR values, where all the bounds are derived using the S2 pulse. Meanwhile, (\ref{eqn:LB-AP-BLOIC}) is given as a benchmark since it can be viewed as the case of $r=\infty$, noting that as $r \to \infty$, $\mu$ tends to $r$, and the RHS of (\ref{eqn:LB-CPAPR-BLOIC}) monotonically increases and tends to the RHS of (\ref{eqn:LB-AP-BLOIC}). For our bounds, it is shown that the PAPR constraint only causes some SNR loss. Moreover, an example of the capacity lower bound of the BLOIC with input constraint $\mathcal P_{\{X\}}\leq \mathcal{E},\mspace{4mu} 0\leq x(t)\leq \mathcal{A}_0$  (called AP-PP-BLOIC, where $\mathcal{A}_0$ is a constant) is given, denoted as AP-PP LB. It is obtained by setting PAPR $r=\mathcal A_0/\mathcal E$ in Corollary 3 for each SNR. The value of $\mathcal{A}_0$ is set to be 10 dB higher than the noise variance. When the SNR is relatively low, the PAPR is large so that the bound is close to the Exp-S2 LB. When the SNR exceeds 7 dB the bound stops increasing.

\section{Discussions and Conjectures}

This section discusses further improvement of our lower and upper bounds, since there are still considerable gaps between them. We believe that new bounding techniques are needed to tighten the gap. We give two conjectures which considers lower and upper bounding, respectively.

\begin{figure}
\centering
\includegraphics[width=3.8in,height=2.85in]{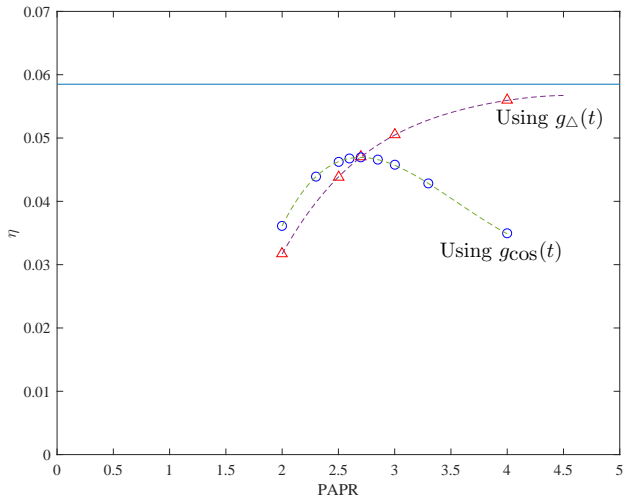}
\caption{Behavior of $\eta$ for two specific modulation pulses in the PAPR-BLOIC.}
\label{fig:beta}
\end{figure}

Our DOF-optimal capacity lower bounds have a general form of
\begin{equation}
\label{eqn:beta}
\mathcal C^\textrm{PC}_\textrm{BLOIC}\ge W\log \left( 1+ \eta\frac{\mathcal {E}^2}{N_0W}\right).
\end{equation}
In Theorem 2 and Theorem 3, $\eta$ can be maximized by finding out the optimal $g(t)$ that maximizes $\mathcal G /\mathcal S_\textrm N ^2$. However, this is still an open problem. Moreover, the optimal $g(t)$ for the PAPR-BLOIC may vary for different PAPR values. In Fig. \ref{fig:beta}, the values of $\eta$ in the lower bounds on the capacity of the PAPR-BLOIC obtained by using $g_\triangle (t)$ and $g_{\textrm {cos}}(t)$ in Theorem 3 are given for different PAPR values. It is shown that when $r$ is smaller than 2.7, $g_{\textrm {cos}}(t)$ is better, and otherwise $g_\triangle (t)$ is better. For large PAPR values, finding out a pulse which achieves larger $\eta$ than that obtained by $g_\triangle(t)$ is difficult, because a $G(f)$ with relatively large $\mathcal G$ always has large sidelobes in the time domain, which also causes large $\mathcal S$. Due to the time-frequency uncertainty, we cannot make $\mathcal G /\mathcal S_\textrm N^2$ very large. In summary, we have the following conjecture:

\emph{Conjecture 1}: The high-SNR asymptotic capacity expression of i.i.d. Nyquist rate PAM signaling over the AP-BLOIC, denoted as $\mathcal C_{\textrm {BLOIC}}^{\textrm{AP, i.i.d. NRP}}$, satisfies
\begin{equation}
\lim\limits_{\mathrm{SNR}\to\infty} \left\{\mathcal C_{\textrm {BLOIC}}^{\textrm{AP, i.i.d. NRP}}-W\log \left(1+\frac{1}{2\pi e}\frac{{\mathcal E}^2}{N_0 W}\right)\right\}=0.
\end{equation}

For the capacity of the AP-DTOIC, the high-SNR asymptotically tight bounds reported in \cite{HK04} (implicitly) and in \cite{LMW09, FH10} imply that
\begin{equation}
\label{eqn:C-AP-DTOIC-asymptotic}
\lim \limits_ {\mathrm{SNR}\to\infty}\left\{ \mathcal C_{\textrm{DTOIC}}^{\textrm{AP}}-\log \frac{\lambda {\cal E}}{\sigma }\right\}=0
\end{equation}
where $\lambda=\sqrt {\frac{e}{2\pi }}$. Our tightest capacity bounds for the AP-BLOIC at high SNR, (\ref{eqn:LB-AP-BLOIC}) and (\ref{eqn:UB1-AP-BLOIC}), have the same pre-log factor but different pre-SNR factors:
\begin{equation}
\label{eqn:Bound-AP-BLOIC}
2W\log\left(\lambda \frac{\cal E}{\sqrt{N_0W}}+2\lambda \right)
\ge \mathcal C_{\textrm{BLOIC}}^{\textrm {AP}} \ge 2W\log \frac{\lambda}{e}\frac{\cal E}{\sqrt{N_0W}}.
\end{equation}
We may thus expect that
\begin{equation}
\label{eqn:C-AP-BLOIC-asymptotic}
\lim \limits_ {\mathrm{SNR}\to\infty}\left\{ \mathcal C_{\textrm{BLOIC}}^{\textrm{AP}}- 2W\log \frac{\rho \lambda {\cal E}}{\sqrt{N_{0}W}}\right\}=0.
\end{equation}
where $\rho \leq 1$ is a factor determining the high SNR capacity of the AP-BLOIC.
The existence and value of the factor $\rho $ in (\ref{eqn:C-AP-BLOIC-asymptotic}) is of interest to us because if $\rho$ is strictly less than one, then we can conclude that when we simplify the BLOIC to a DTOIC which transmits $2W$ times per second, there does exist a penalty on capacity because of the fundamental distinction between the BLOIC and the DTOIC. We have the following conjecture.

\emph{Conjecture 2}: The factor $\rho $ in (\ref{eqn:C-AP-BLOIC-asymptotic}) exists and satisfies $\rho <1$.

\emph{Remark 2:} A possible way of settling Conjecture 2 is using the sphere packing based upper bounding technique in \cite{YK02}, by which we can get a high-SNR asymptotic upper bound on the capacity of the AP-BLOIC as
\begin{equation}
\label{eqn:UB-asymptotic-AP-BLOIC}
\begin{split}
\lim \limits_{\mathrm{SNR}\to\infty} &\bigg\{ \mathcal C_{\textrm{BLOIC}}^{\textrm{AP}}\\
&-W\log\left(\lim \limits_{K\to\infty}K\left(\mathrm {Vol}(\Upsilon^K) \right)^{\frac{1}{K}}\frac{2}{\pi e}\frac{\mathcal{E}^{2}}{N_{0}W}\right)\bigg\}\leq0,
\end{split}
\end{equation}
where $\Upsilon^K$ is the admissible region of length-$K$ input symbol sequences as
\begin{align}
\label{eqn:AdmissibleRegion}
&\Upsilon^K=\notag\\
&\bigg\{[c_k]_{k=1}^{K}:\frac{1}{2}+\mathrm {Re}\left[\sum\limits_{k=1}^{K}c_k\exp \frac{-j2k\pi Wt}{K}\right]\ge 0, \mspace{4mu}c_k \in \mathbb C\bigg\},
\end{align}
and $\mathrm {Vol}(\Upsilon^K)$ is its volume. Conjecture 2 can be proved if we can show that $\lim \limits_{K\to\infty}K\left(\mathrm {Vol}(\Upsilon^K) \right)^{\frac{1}{K}} <e^2/4$. However, a direct calculation of this limit or even its upper bound is nontrivial. In \cite{YK02} it was proved that $\Upsilon^K$ is a subset of a $K$-dimensional trigonometric moment space $\mathcal M^K$ whose volume has been determined to satisfy $\lim \limits_{K\to\infty}K\left(\mathrm {Vol}(\mathcal M^K) \right)^{\frac{1}{K}} <\pi e/2$. Thus we have the asymptotic upper bound shown in Fig. \ref{fig:BLOIC-basic}, based on (\ref{eqn:UB-asymptotic-AP-BLOIC}). Unfortunately this is not enough for settling Conjecture 2.

\emph{Note:} In fact, $\Upsilon^K$ is the admissible region of the input of the AP-MCOIC with $K$ subcarriers and a nominal bandwidth $W$. As $K\to \infty$, \cite{YK02} shows that the high-SNR asymptotic capacity of the AP-MCOIC is upper bounded by $W\log\frac{\mathcal{E}^2}{N_{0}W}$.\footnote{The expression of this result is different from [\ref{bYK02}, (56)] (when the channel gain is normalized to one) because \cite{YK02} used a nonstandard definition of the power spectral density of the white Gaussian noise.} We note that each asymptotic result obtained by (\ref{eqn:UB-asymptotic-AP-BLOIC}) is also a high-SNR asymptotic upper bound on the capacity of the AP-BLOIC, although it is obtained from considering the AP-MCOIC. The interpretation is as follows. For a fixed $W$, $K$ tending to infinity is equivalent to the length of a $\frac{1}{2W}$-interval sample sequence of a block of input of the MCOIC tending to infinity. Meanwhile, the out-of-band energy of the MCOIC in the sense of nominal bandwidth decreases to zero, and the time domain $\frac{1}{2W}$-interval samples of the MCOIC reduce to the Nyquist samples. In summary, as $K$ tends to infinity, the limit of the admissible region of the input of the AP-MCOIC tends to the admissible region of the input of the AP-BLOIC with bandwidth $W$. So the capacity of the AP-MCOIC converges to the capacity of the AP-BLOIC whose bandwidth is equal to $W$.

\begin{appendices}
      \section{  }

Consider an i.i.d. Nyquist rate PAM ensemble using $g_\triangle(t)$ as
\begin{align}
\label{eqn:Xtriangular}
X_\triangle(t)&=\sum_{i}X_{i}g_\triangle\left(t-\frac{i}{2W}\right)\notag\\
&=\sum \limits _{i}X_i \frac{\sin ^2\pi W(t-i/2W)}{2(\pi W(t-i/2W))^2}.
\end{align}
Let the input symbols be i.i.d. nonnegative (which guarantees the nonnegativity of $X_\triangle(t)$ since $g_\triangle(t)$ is nonnegative) and let $\mathrm{E}[X]$ be equal to $\mathcal{E}$. Then the AP of $X_\triangle(t)$ is
\begin{align}
\label{eqn:P=E}
 &\mathcal P_{\{X_\triangle\}}\notag\\
 &=\mathop {\lim }\limits_{T\to \infty } \frac {1}{2T} \mathrm E\left[\int_{-T}^{T} {\sum\limits_{i=-\infty}^{\infty} {X_i \frac{\sin ^2\pi W(t-iT_\textrm{s} )}{2\left( {\pi W(t-iT_\textrm{s} )} \right)^2}}\mathrm{d}t}\right]\notag\\
 &=\mathop {\lim }\limits_{N\to \infty } \frac {1}{2NT_\textrm s} \int_{-NT_\textrm s}^{NT_\textrm s}{ {\sum\limits_{i=-\infty}^{\infty} {\mathrm E\left[X_i \right]\frac{\sin ^2\pi W(t-iT_\textrm{s} )}{2\left( {\pi W(t-iT_\textrm{s} )} \right)^2}}}\mathrm{d}t}\notag\\
 &=\mathcal E\cdot \mathop {\lim }\limits_{N\to \infty } \frac {1}{2NT_\textrm s}\sum\limits_{n=-N}^{N-1} \int_{nT_\textrm s}^{(n+1)T_\textrm s}{ {\sum\limits_{i=-\infty}^{\infty} {\frac{\sin ^2\pi W(t-iT_\textrm{s} )}{2\left( {\pi W(t-iT_\textrm{s} )} \right)^2}}}\mathrm{d}t}\notag\\
 &=\mathcal E\cdot\mathop {\lim }\limits_{N\to \infty } \frac {1}{2NT_\textrm s} \sum\limits_{n=-N}^{N-1} { {\sum\limits_{i=-\infty}^{\infty}\int_{nT_\textrm s}^{(n+1)T_\textrm s}{{ \frac{\sin ^2\pi W(t-iT_\textrm{s} )}{2\left( {\pi W(t-iT_\textrm{s} )} \right)^2}\mathrm{d}t}}}}\notag\\
 &=\mathcal E\cdot\mathop {\lim }\limits_{N\to \infty } \frac {1}{2NT_\textrm s} \sum\limits_{n=-N}^{N-1} { {\sum\limits_{i=-\infty}^{\infty}\int_{(n-i)T_\textrm s}^{(n+1-i)T_\textrm s}{{ \frac{\sin ^2\pi Wt'}{2\left( {\pi Wt'} \right)^2}\mathrm{d}t'}}}}\notag\\
 &=\mathcal E\cdot\mathop {\lim }\limits_{N\to \infty } \frac {1}{2NT_\textrm s}2N\int_{-\infty}^{\infty} { {\frac{\sin ^2\pi Wt}{2\left({\pi Wt} \right)^2}}\mathrm{d}t}\notag\\
 &=\mathcal E\cdot\frac{1}{T_\textrm{s}}\cdot\frac{1}{2W}\notag\\
 &=\mathcal E,
\end{align}
where the second equality follows by the linearity of expectation and letting $T=NT_\textrm s$, the third by dividing the interval of integration, the fourth by noting that the infinite sum converges for $t\in \mathbb{R}$, the fifth by defining $t'\triangleq t -iT_\textrm s$, and the sixth by noting that the integrals over all the intervals $[(n-i)T_\textrm s, (n+1-i) T_\textrm s)$ sum to the integral over $(-\infty, \infty)$.
Thus the AP constraint is satisfied, and $X_\triangle(t)$ is admissible for the AP-BLOIC. Since
\begin{equation}
\label{eqn:Ftriangular}
G_\triangle(f)=
\begin{cases}
\frac{1}{2W} \frac{W-|f|}{W},&|f|\leq W \\
 0,&\textrm {otherwise,}
 \end{cases}
\end{equation}
we have
\begin{align}
\label{Gtri}
\mathcal G&=\exp\left(\frac{1}{W}\int_{0}^W {\log \left| {2W\cdot G_\triangle (f)} \right|^2\mathrm{d}f} \right)\notag\\
&=\exp\left(\frac{2}{W}\int_0^W {\log \left({\frac{W-f}{W}} \right)\mathrm{d}f}\right)\notag\\
&=\frac{1}{e^2}.
\end{align}
Substituting (\ref{Gtri}) into (\ref{eqn:LB-PAM}) completes the proof of Theorem 1.

      \section{  }
Consider the i.i.d. Nyquist rate PAM ensemble $X_\textrm{PAM}(t)$ as (\ref{eqn:PAM-Nyquist}). By letting $X_i$ be bounded within $[-\frac{\cal A}{2 \mathcal S_\textrm N},\frac{\cal A}{2 \mathcal S_\textrm N}]$, we can make the ensemble satisfy $-\frac{\cal A}{2}\leq X_\textrm{PAM}(t)\leq \frac{\cal A}{2}$, according to the definition of $\mathcal S_\textrm N$. Let a DC bias $\mathcal D=\mathcal A/2$ be added to $X_\textrm{PAM}(t)$, and thus the nonnegativity and PP constraints of the PP-BLOIC are both satisfied. The maximum differential entropy of $X_i$ is $h(X)=\log\frac {\mathcal A}{\mathcal S_\textrm N}$ and is obtained by letting $X_i$ be uniformly distributed. Using Lemma 2, (\ref{eqn:LB-Shamai-General-PP-BLOIC}) is obtained immediately.

      \section{  }

For the case of $r>2$, let us consider the i.i.d. Nyquist rate PAM ensemble $X_\textrm{PAM}(t)$ as (\ref{eqn:PAM-Nyquist}). Assume the inputs $X_i$ to be bounded within $[0,L]$ and let $\mathrm E [X_i]=L/\nu $, $\nu\ge 2$, where $\nu$ is the PAPR of $X_i$. The maxentropic distribution of $X_i$ is a truncated exponential distribution according to [\ref{bLMW09}, eqn. (42)],
\begin{equation}
\label{TEPDF}
p_X(x)=\frac{1}{L}\frac{\mu}{1-e^{-\mu}}e^{-\frac{\mu x}{L}},\mspace{10mu} 0\leq x\leq L,
\end{equation}
where $\mu$ is the unique solution of
\begin{equation}
\frac{1}{\nu}=\frac {1}{\mu}-\frac {e^{-\mu}}{1-e^{-\mu}}.
\end{equation}
The differential entropy of (\ref{TEPDF}) is
\begin{equation}
\label{TEH}
h(X)=\frac {\mu}{\nu} + \log \left(L\frac{1-e^{-\mu}}{\mu}\right).
\end{equation}
Then the maximum achievable information rate of $X_\textrm{PAM}(t)$ can be lower bounded by
\begin{equation}
\label{LB-APPP-General-1}
\mathcal R[X(t)]\ge W\log \left( 1+ \frac { \mathcal G \nu^2 e^{2\mu/\nu}}{2\pi e }\left( \frac{1-e^{-\mu}}{\mu}\right)^2 \frac {L^2}{\nu ^2 N_0W}\right),
\end{equation}
which follows from Lemma 2 by using the distribution (\ref{TEPDF}).

We now design $X_\textrm{PAM}(t)$ to be an admissible input ensemble of the PAPR-BLOIC with PAPR $r$ and convert (\ref{LB-APPP-General-1}) into a capacity lower bound of the PAPR-BLOIC with parameters $r$, $\mathcal E$, $\mathcal G$, and $\mathcal S_\textrm N$. 
We first shift the distribution of $X_i$ by $-\frac{L}{2}$ so that $X_i$ is distributed in $\left[-\frac{L}{2}, \frac{L}{2}\right]$. Now its mean becomes
\begin{equation}
\mathrm E[X_i]=\frac{L}{\nu}-\frac{L}{2}.
\end{equation}
The corresponding i.i.d. Nyquist rate PAM ensemble, denoted as $X_0(t)$, satisfies
\begin{equation}
-\frac{\mathcal S_\textrm N L}{2}\leq X_0(t)\leq \frac{\mathcal S_\textrm N L}{2}.
\end{equation}
By adding a DC bias $\mathcal D=\frac{\mathcal S_\textrm N L}{2}$ on $X_0(t)$, we obtain an admissible waveform ensemble $X_\textrm{PAM}(t)$ that satisfies $0\leq X(t)\leq \mathcal S_\textrm N L$. The PP constraint is satisfied by letting $\mathcal S_\textrm N L$ be equal to $r\mathcal E$. Since the mean of $X_0(t)$ is equal to $\mathrm E[X_i]$, i.e. the mean of the input symbol (cf. the derivation of (\ref{eqn:P=E})), the AP of $X_\textrm{PAM}(t)$ is
\begin{equation}
\label{APXt}
\mathcal E=\mathcal D+\mathrm E[X_i] = \frac{2-\nu +\nu \mathcal S_\textrm N}{2\nu}L.
\end{equation}
The PAPR of $X_\textrm{PAM}(t)$ is $r$ if we set $\nu$ as
\begin{equation}
\label{nu}
\nu=\frac{2 r}{2\mathcal S_\textrm N-r\mathcal S_\textrm N+ r}.
\end{equation}
According to (\ref{nu}) and $\mathcal S_\textrm N L=r\mathcal E$, we can convert (\ref{LB-APPP-General-1}) into a general capacity lower bound of the PAPR-BLOIC as the case of $r>2$ in Theorem 3.

For the case of $0<r\leq2$, consider the designed input ensemble $X_\textrm{PAM}(t)$ (after DC bias being added) in the proof of Theorem 2 in Appendix B. Apparently the AP of $X_\textrm{PAM}(t)$ is $\mathcal A/2$. If we let $\mathcal A=r\mathcal E$, the AP of $X_\textrm{PAM}(t)$ is then $r\mathcal E/2$ which is smaller than $\mathcal E$ since $r<2$. So $X_\textrm{PAM}(t)$ is also admissible for the PAPR-BLOIC with $r<2$ and $r\mathcal E=\mathcal A$. Replacing $\mathcal A$ in (\ref{eqn:LB-Shamai-General-PP-BLOIC}) by $r\mathcal E$, the case of $0<r<2$ is obtained and the proof of Theorem 3 is completed.

Corollary 3 follows immediately from Theorem 3 by letting $g(t)$ be $g_\triangle (t)$ and noting that in this case $\mathcal G=1/e^2$ (see (\ref{Gtri})) and $\mathcal S_\textrm N=1$ (this was implicitly shown in \cite{Shannon48}).

      \section{  }

Consider the following input ensemble which achieves ISI-free signaling according to \cite{H07}:
\begin{equation}
\label{eqn:Xtriangular-ISI-free}
X_\triangle^{\textrm{IFS}}(t)=\sum \limits _{i} X_i \frac{\sin ^2\pi W(t-i/W)}{(\pi W(t-i/W))^2},
\end{equation}
where the pulse used is a scaling of $g_\triangle (t)$ with a factor of two so as to satisfy Definition 1. Letting the input symbols $\{X_i\}$ be i.i.d. and satisfy $\mathrm E[X]=\mathcal E$, then the AP of the input ensemble is $\mathcal P_{\{X\}}=\mathcal E$ (cf. the derivation of (\ref{eqn:P=E}), noting that the time domain integral of the pulse used here is $1/W$). Denote the equivalent discrete-time memoryless channel of ISI-free signaling over the BLOIC using (\ref{eqn:Xtriangular-ISI-free}), which is obtained from $\frac 1W$-interval sampling at the receiver, as
\begin{equation}
\label{eqn:DTequivalent}
Y[i]=X[i]+Z[i].
\end{equation}
The value of $g(0)$ determines the noiseless sample values when the corresponding input symbols are given. Since $2\cdot g_\triangle (0)=1$, the noiseless samples satisfy $X[i]\equiv X_i$. The variance of the noise samples can be minimized by an ideal bandlimited filtering over $\mathcal{W}$ on the received noisy waveform, not affecting the value of $X[i]$ in (\ref{eqn:DTequivalent}) and preserving the memoryless property of noise samples. So the obtained noise samples $Z[i]$ are i.i.d. with variance $N_{0}W$. Then if we let $X_i$ be exponentially distributed as (\ref{eqn:EPDF}), we have
\begin{align}
\label{eqn:LB-eqivalentISI}
\mathcal R[X_\triangle^{\textrm{IFS}}(t)]
 &=W\cdot I(X[i], Y[i])\notag\\
 &=W\cdot( h(Y[i])-h(Z[i]))\notag\\
 &\ge \frac{W}{2}\log\left(1+e^{2h(X_i)-2h(Z[i])}\right)\notag\\
 &=\frac{W}{2}\log\left(1+\frac{e}{2\pi}\frac{{\mathcal E}^2}{N_0 W}\right),
\end{align}
where the inequality follows from the EPI, and (\ref{eqn:LB-ISI}) is obtained.

The proofs of (\ref{eqn:LB-ISI-U-1}), the first case of (\ref{eqn:LB-ISI-U}), and the second case of (\ref{eqn:LB-ISI-U}) are similar except that we let $X_i$ be uniformly distributed in $[0,\mathcal A]$, truncated-exponentially distributed in $[0,r\mathcal E]$ (whose PDF can be obtained from (\ref{TEPDF}) by replacing $L$ with $r\mathcal E$ and replacing $\nu$ with $r$), and uniformly distributed in $[0,r\mathcal E]$, respectively.

      \section{  }

The proof of Theorem 5 can be done by a variation on the proof of Theorem 4 as follows. Note that (\ref{eqn:DTequivalent}) can be viewed as a DTOIC where $Z_i\sim \textrm N(0, N_0W)$, transmitting only $W$ times per second. So we have
\begin{equation}
\label{eqn:ISItoDTOIC}
\mathcal R_{\textrm{BLOIC}}^{\textrm{AP},\mspace{4mu}\textrm {IFS}}\ge \mathcal C_{\textrm{DTOIC}}^{\textrm{AP},\mspace{4mu} \sigma^2=N_0W}\cdot W\mspace{10mu}\textrm {transmissions per second}.
\end{equation}
Now all capacity lower bounds of the AP-DTOIC can be used to derive a lower bound for the maximum achievable information rate of ISI-free signaling over the AP-BLOIC. The tightest known capacity lower bound of the DTOIC is in \cite{FH10}, where a geometric distribution as (\ref{Geo}) is used as the input distribution. The parameter $l$ (space between mass points of geometric distribution) in (\ref{Geo}) is optimized \emph{for each SNR} to maximize the information rate achieved by such lower bound. That information rate is thus $\max\limits_l I(\textsf Q_\textrm{g}(l), \textsf V)$. The proof of Theorem 5 is then completed by replacing $\mathcal C_{\textrm{DTOIC}}^{\textrm{AP},\mspace{4mu} \sigma^2=N_0W}$ in (\ref{eqn:ISItoDTOIC}) by $\max\limits_l I(\textsf Q_\textrm{g}(l), \textsf V)$.

      \section{  }

Consider the i.i.d PAM ensemble $X_\textrm{PAM}(t)$ as (\ref{eqn:PAM}) in which we let $g(t)$ be a Nyquist pulse satisfying Definition 1 and $T_\textrm s$ be $\frac{1+\beta}{2W}$ so that the ISI-free property is achieved. Let the input symbols $X_i$ be uniformly distributed in $[-L/2, L/2]$. Since $\mathrm E[X_i]=0$, the mean of the ensemble obtained is also zero (cf. the derivation in (\ref{eqn:P=E})). According to the definition of $\mathcal S(\tau)$, we have $-\frac{1}{2}\mathcal S_\beta L\leq X_\textrm{PAM}(t)\leq \frac{1}{2}\mathcal S_\beta L$. So the DC bias needed is $\frac{1}{2}\mathcal S_\beta L$, and the mean (AP) of $X_\textrm{PAM}(t)$ becomes
\begin{equation}
\label{ESA}
\mathcal E=\frac{\mathcal S_\beta L}{2},
\end{equation}
with that DC bias.

Consider the equivalent discrete-time channel of such ISI-free signaling, denoted as $Y[i]=X[i]+Z[i]$ (cf. the derivation of (\ref{eqn:LB-eqivalentISI})). Since $X[i]$ is uniformly distributed, from (\ref{ESA}) we have
\begin{equation}
h(X[i])=\log L=\log \frac {2\mathcal E}{\mathcal S_\beta}.
\end{equation}
So the maximum achievable information rate can be lower bounded as (cf. (\ref{eqn:LB-eqivalentISI}))
\begin{align}
\label{RPL}
\mathcal R
 &=\frac{2W}{1+\beta}\cdot I(X[i], Y[i])\notag\\
 &\ge \frac{W}{1+\beta}\log\left(1+e^{2h(X[i])-2h(Z[i])}\right)\notag\\
 &=\frac{W}{1+\beta}\log \left(1+\frac{2}{\mathcal S_\beta^2\pi e} \frac{\mathcal E^2}{N_0W} \right),
\end{align}
where the inequality follows from the EPI, and (\ref{LB-ISI-bias}) is obtained immediately. This completes the proof of (\ref{LB-ISI-bias}).

Since the input ensemble obtained in the proof of (\ref{LB-ISI-bias}) is bounded in $[0, \mathcal S_\beta L]$, by letting $L=\frac{\mathcal A}{\mathcal S_\beta}$ the same input ensemble is admissible for the PP-BLOIC. Combining $L=\frac{\mathcal A}{\mathcal S_\beta}$, (\ref{ESA}), and (\ref{RPL}) we obtain (\ref{LB-ISI-bias-1}).

Consider a truncated exponential distribution $\textsf Q_{\textrm {TE}}$ whose PDF is as (\ref{TEPDF}) except letting
\begin{equation}
\label{TEcondition}
\mathcal S_\beta L=r\mathcal E,\mspace{10mu} \nu=\frac{2r}{2\mathcal S_\beta-r\mathcal S_\beta+r}.
\end{equation}
From the proof of Theorem 3, we know that if we use such an input symbol distribution to replace the uniform distribution used in the proof of (\ref{LB-ISI-bias}), the obtained input ensemble is admissible for the PAPR-BLOIC (i.e. the AP equals $\mathcal E$ and the PAPR equals $r$) after adding a minimum required DC bias. Replace the $h(X[i])$ in (\ref{RPL}) by the differential entropy of $\textsf Q_{\textrm {TE}}$ (which is obtained by combining (\ref{TEcondition}) and (\ref{TEH})) we obtain the first case of (\ref{LB-ISI-bias-2}). For the second case we alternatively use an uniform distribution $\textsf Q_{\textrm {U}}$ in $\left[0,\frac{ r\mathcal E}{\mathcal S_\beta}\right]$ which guarantees the obtained input ensemble satisfying the PP constraint. Noting that the AP constraint is also satisfied, the second case of (\ref{LB-ISI-bias-2}) can be obtained from replacing the $h(X[i])$ in (\ref{RPL}) by the differential entropy of $\textsf Q_{\textrm {U}}$. This completes the proof of Theorem 6.

      \section{  }

Consider a bandlimited channel as $Y(t)=X(t)+Z(t)$ (where $Z(t)$ is defined as in (\ref{eqn:BLOIC})), denoted as channel A, with a relaxed version of input power constraint $\textrm {PC}$ as follows: 1) the nonnegativity and PP constraints on $X(t)$ hold only at $t=n/2W, \mspace{4mu} \forall n \in \mathbb{Z}$, i.e. on a sequence of Nyquist sample points; 2) the AP constraint is defined like that in the BLOIC. Obviously, the capacity of this channel, denoted as $\mathcal C_\textrm{A}^\textrm {PC-relaxed}$, is an upper bound of $\mathcal C_\textrm{BLOIC}^\textrm{PC}$. Each input ensemble of channel A, denoted as $X_\textrm{A}(t)$, can be viewed as an i.i.d. Nyquist rate PAM ensemble as (\ref{eqn:PAM-Nyquist}) where $g(t)=g_\textrm{sinc}(t)$ and the input symbols $\{X_i\}$ are just the Nyquist samples of $X_\textrm{A}(t)$. By matched filtering $X_\textrm{A}(t)$ with $2W\cdot G_\textrm{sinc}(f)$ and sampling at Nyquist intervals (an information lossless procedure), an equivalent model of channel A is obtained as the following discrete-time memoryless channel:
\begin{equation}
Y[i]=X[i]+Z[i],
\end{equation}
where $X[i] =X_i$, and $Z[i] \sim \textrm N(0, N_0W)$. This channel is a DTOIC with the same type of constraints and corresponding parameters per the constraint $\textrm {PC}$ of channel A. Since this channel transmits $2W$ symbols per second, we have
\begin{align}
\label{eqn:BLDTrelation2}
\mathcal C_{\textrm{BLOIC}}^{\textrm{PC}} &\le \mathcal C_{\textrm{A}}^{\textrm{PC-relaxed}}\notag\\
&=\mathcal C_{\textrm{DTOIC}}^{\textrm{PC}, \sigma^2=N_0W}\cdot 2W\mspace{8mu}\textrm {transmissions per second}.
\end{align}
This completes the proof of Lemma 3.

  \end{appendices}

\section*{Acknowledgements}

The authors would like to thank the anonymous reviewers for their comments which helped to significantly improve the manuscript.

\ifCLASSOPTIONcaptionsoff
  \newpage
\fi

\end{document}